\documentclass{jps-cp}
\usepackage{txfonts} 
\usepackage{subfigure}
\usepackage{url}

\usepackage{color}

\title{Exclusive Hard Processes for Studying Hadron Structures at J-PARC}

\author{Wen-Chen \textsc{Chang}$^{1}$\thanks{Speaker}}

\inst{$^{1}$Institute of Physics, Academia Sinica, Taipei 11529,
  Taiwan}

\email{changwc@phys.sinica.edu.tw}

\recdate{December 23, 2019}

\abst{With an appropriate hard scale, exclusive hadronic processes
  could provide novel information of the internal quark-gluon
  configurations of hadrons. The availability of 10-20 GeV secondary
  meson beam in the coming high-momentum beam line of Hadron Hall at
  J-PARC offers an unique opportunity to carry out the measurements of
  the exclusive hard processes. We address this interesting approach
  by two possibilities: (a) $\pi^- p \to K^0 \Lambda(1405)$ for the
  constituent quark structure of $\Lambda(1405)$, and (b) exclusive
  pion-induced Drell-Yan process $\pi^- p \to \gamma^* n \to l^+ l^-
  n$ for the generalized parton distributions (GPDs) of
  nucleons. Realization of such measurements at J-PARC will open up a
  new way of accessing the internal quark structure of exotic hadrons
  and also the nucleon GPDs based on a solid theoretical foundation of
  perturbative QCD.}

\kword{Exclusive hard process, Hadron Structure, $\Lambda(1405)$,
  Drell-Yan process}

\begin{document}
\maketitle

\section{Introduction}

Exclusive hadronic processes with a full identification of the initial
and final states are mostly studied at low energies considering large
enough cross sections. Nevertheless, the sensitivity of the reactions
to the partonic structures of hadrons involved is constrained by the
soft energy scale. With the modern high-luminosity accelerators and
large-acceptance detector systems, exclusive processes probed with
hard scales become an effective tools to explore the partonic
structure of hadrons. We consider such possible physics programs in
coming high-momentum beam line of Hadron Hall at J-PARC.

An important feature of the J-PARC accelerator is the high intensity
of the primary 30-GeV proton beam~\cite{J-PARC}. A high-momentum beam
line is under construction at the Hadron Hall now. In addition, by
installing a thin production target at the branching point, one can
obtain unseparated secondary beams such as pions, kaons and
anti-protons. Constrained by the radiation safety of beam loss from a
20-mm long gold target, the maximum intensities of primary proton beam
are $9x10^13$ protons per pulse (ppp) at a 5.2 sec repetition cycle
time (83.8 kW). The momentum profile of secondary beams for $\pi^-$
would be 10-20~GeV with the typical intensity of
$10^{7}$-$10^{8}$/sec. A beam momentum resolution of better than 0.1\%
can be obtained by using the dispersive method. The high momentum
beamline of delivering the primary 30-GeV proton beam is scheduled to
complete in early 2020. Afterward it might take 5 years or more to
make the secondary beam available.

In this proceedings, we consider two kinds of hard exclusive
reactions: exclusive $\Lambda (1405)$ production at large momentum
transfer, and exclusive $\pi$-induced Drell-Yan process. Both of them
are novel approaches to access important partonic structure of hadrons
and are planned to be measured in J-PARC E50 experiment. We introduce
the physic importance of measuring these two processes. The
feasibility of carrying out the exclusive Drell-Yan process will be
discussed in details.

\section{Constituent quark configuration of $\Lambda (1405)$}
\label{sec:measurement}

The exact constituent quark configuration of $\Lambda (1405)$ has been
a long-standing and controversial question. This exotic hadron has
been considered as a conventional 3-$q$ baryon, $\bar{K}N$ molecule or
exotic 5-$q$ pentaquark. There have been tremendous theoretical and
experimental efforts in revealing its true
properties~\cite{Hyodo:2011ur,Mai:2018}. Recently an important result
is the evidence of $\Lambda (1405)$ as $\bar{K}N$ molecule from the lattice
QCD calculation~\cite{Hall:2014uca}. Nevertheless the requirement of
experimental confirmation remains. Despite many measurements of
production and decay channels as well as the medium effect of $\Lambda
(1405)$, there is a lack of conclusive results of its true hadronic
configuration.

The constituent-counting rule based on the perturbative
QCD~\cite{Lepage:1980fj} has been known for a long time. From the
perspective of pQCD, hard gluon exchange should occur to maintain the
exclusive nature of an exclusive hard
process~\cite{Kawamura:2013iia}. In a hard exclusive process of strong
interaction, the degree of power suppression depends on the sum of
constituent-quark degrees of freedom for all participating
particles. The differential cross sections of a hadronic interaction
$a+b \rightarrow c+d$ at a large scattering angle $\theta$ in the CM
system should scale like $d\sigma/dt \sim s^{2-n} f(\theta_{CM})$,
where $s$ and $t$ are Mandelstam variables, $\theta_{CM}$ is the
scattering angle in the CM system and $n$ is the sum of numbers of
elementary constituents in hadron $a$, $b$, $c$ and $d$. The
measurement of scaling factor $n$ of the exclusive production reaction
of the exotic hadron would provide the information of its quark
configuration.

\begin{figure}[htbp]
\begin{center}
\centering
\subfigure[Data of $\gamma \, p \to K^{+} \Lambda$ fit by the scaling factor $n=10.0$.]
{\begin{minipage}[b]{0.49\textwidth} 
 \centering 
  \includegraphics[width=1.0\textwidth]{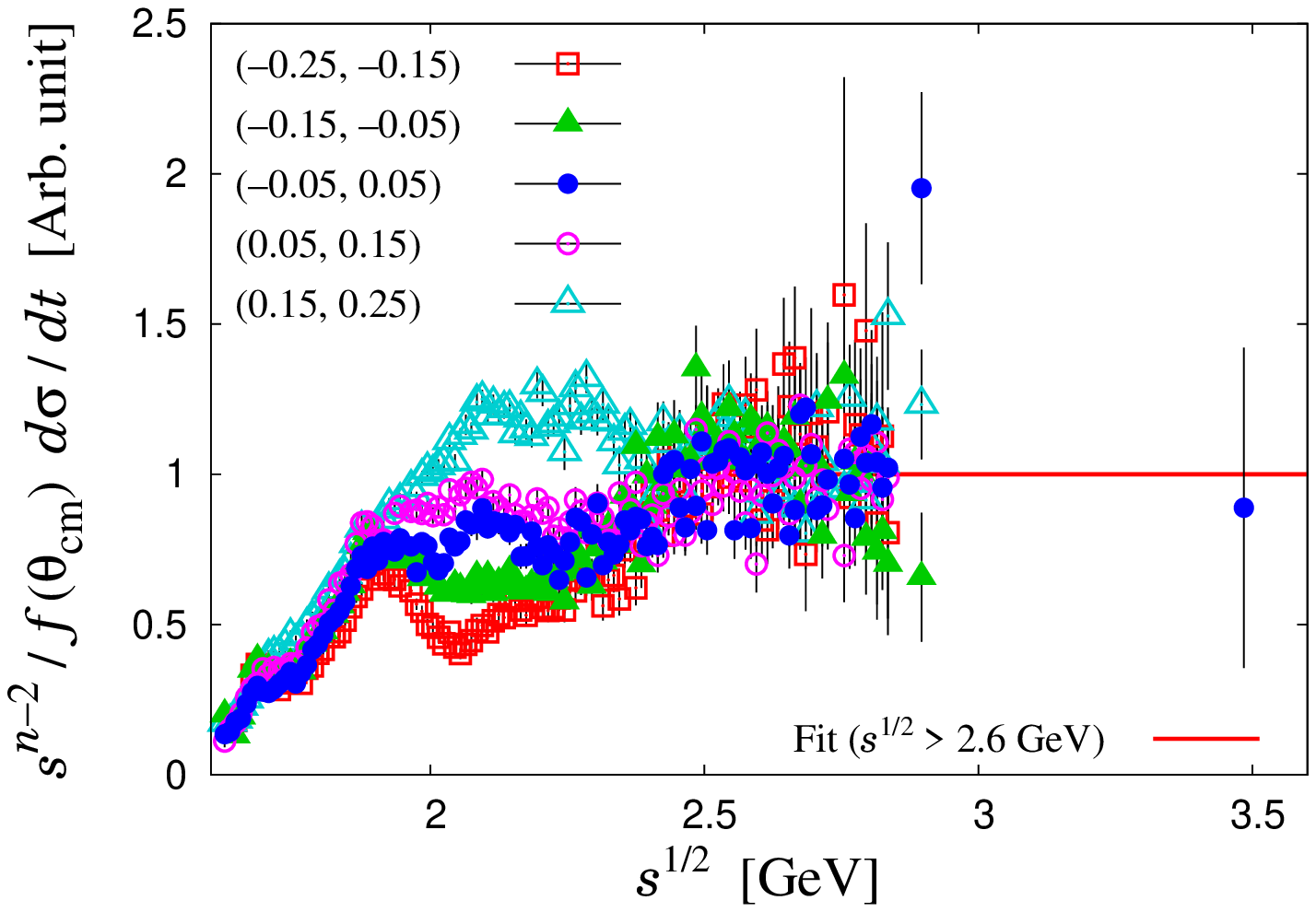}
  \vspace{0.1cm}
  \end{minipage}
\label{fig1a}}
\subfigure[Data of $\gamma \, p \to K^{+} \Sigma ^{0}$ fit by $n=9.2$.]
{\begin{minipage}[b]{0.49\textwidth} 
 \centering 
  \includegraphics[width=1.0\textwidth]{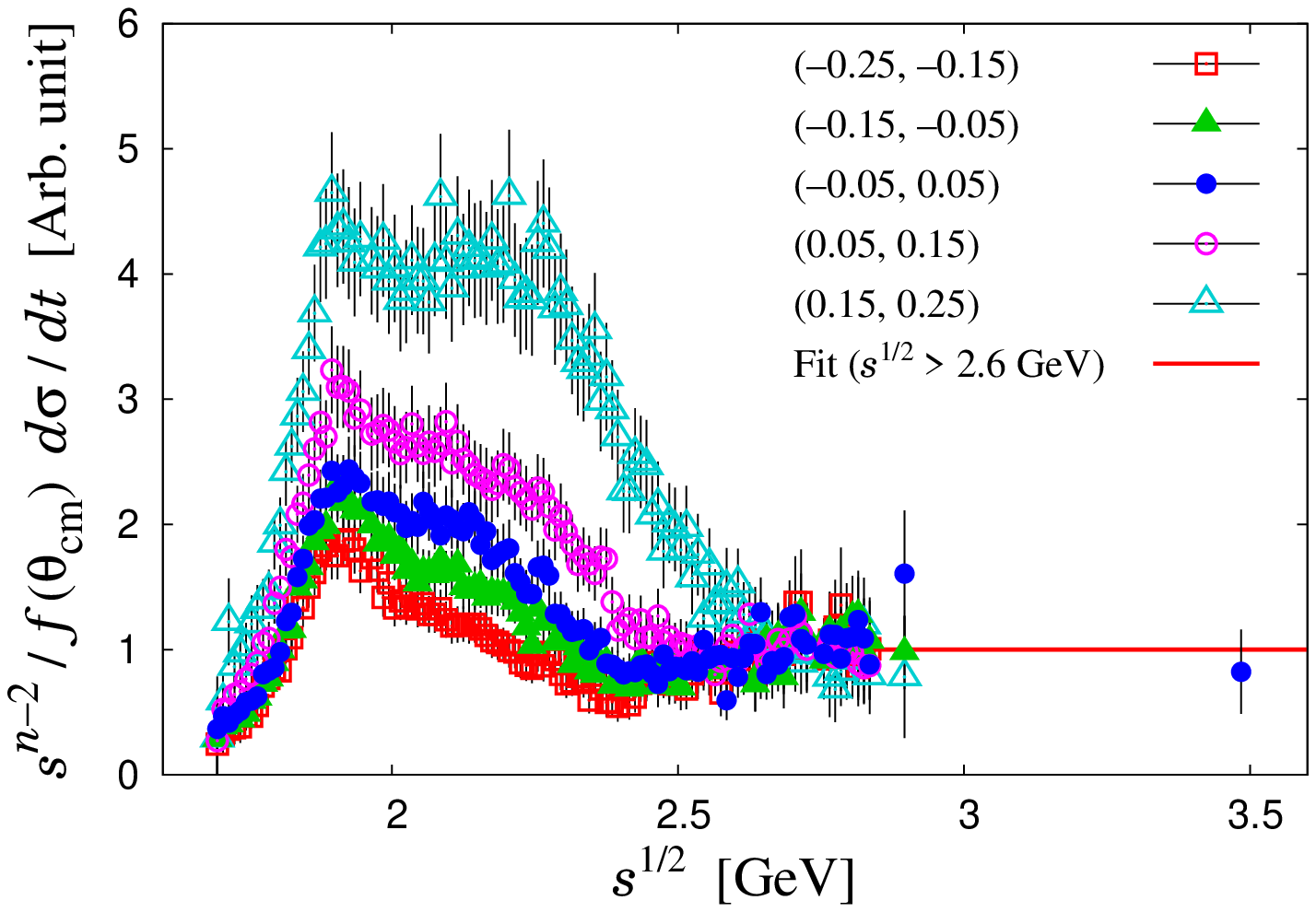}
  \vspace{0.1cm}
  \end{minipage}
\label{fig1b}}
\subfigure[Data of $\gamma \, p \to K^{+} \Lambda (1520)$ fit by $n=9.8$.]
{\begin{minipage}[b]{0.49\textwidth} 
 \centering 
  \includegraphics[width=1.0\textwidth]{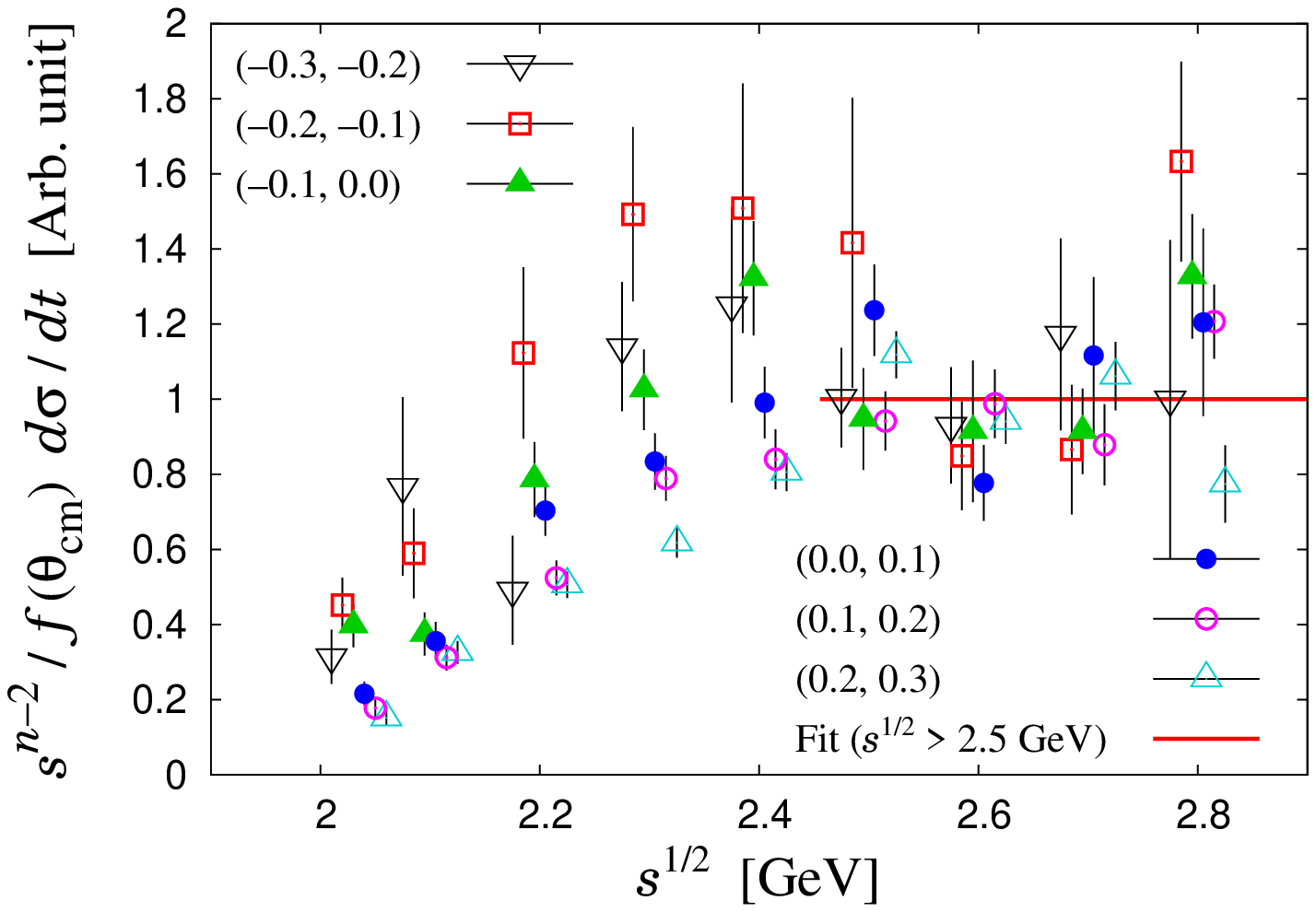}
  \vspace{0.1cm}
  \end{minipage}
\label{fig1c}}
\subfigure[Data of $\gamma \, p \to K^{+} \Lambda (1405)$ fit by $n=10.2$.]
{\begin{minipage}[b]{0.49\textwidth} 
 \centering 
  \includegraphics[width=1.0\textwidth]{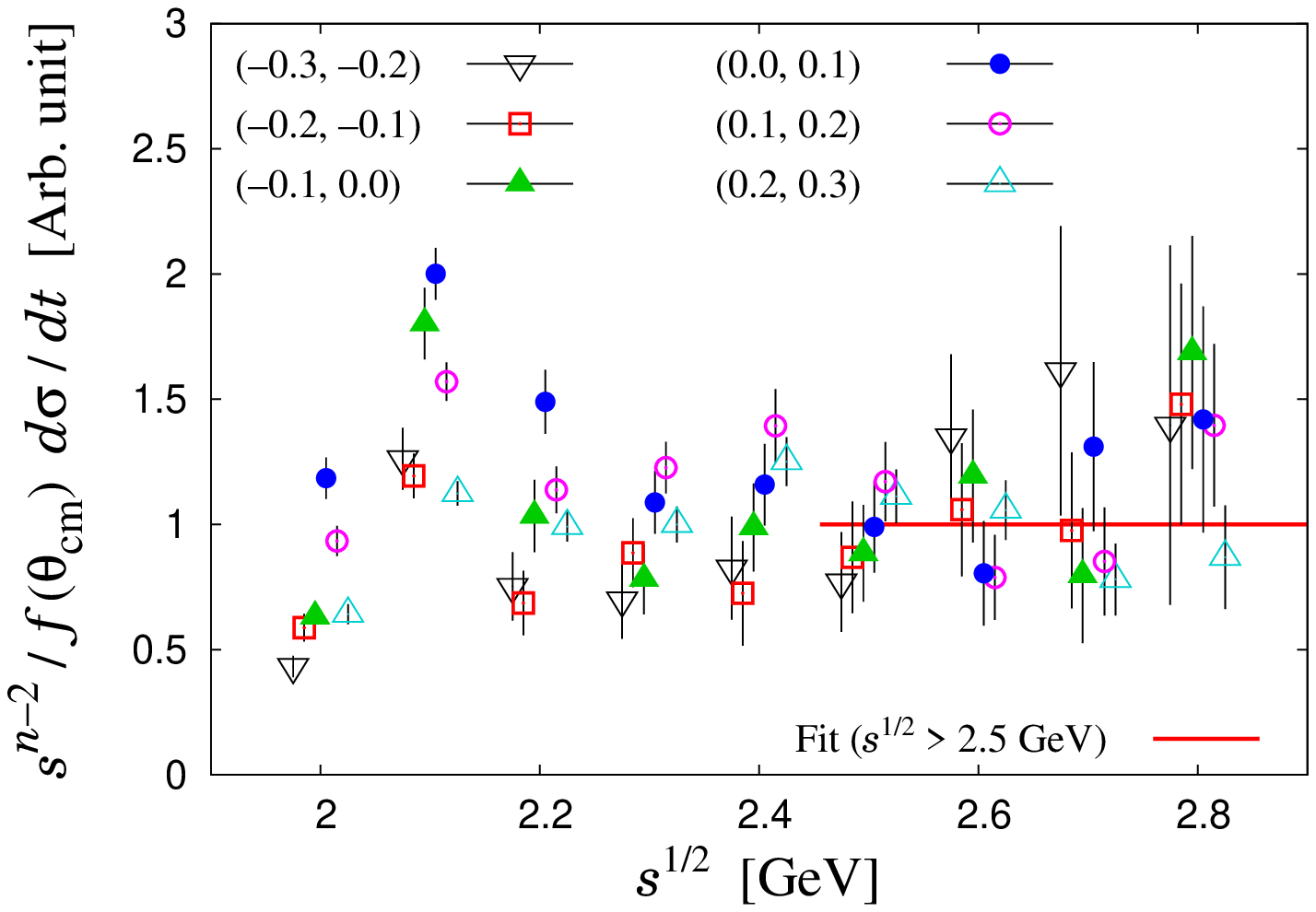}
  \vspace{0.1cm}
  \end{minipage}
\label{fig1d}}
\caption
[\protect{}]{Check of scaling behavior of constituent-counting rule in
  the data of photoproduction of hyperons. Data of $d \sigma/dt$ at
  different bins of $cos \theta_{cm}$ multiplied by
  $s^{n-1}/f(\theta_{cm})$ are denoted by various symbols. An energy
  scaling with the factor of $n=9$ or $n=11$ corresponds to a 3-$q$ or
  5-$q$ nature of the hyperon. Figures from~\cite{Chang:2015ioc}.}
\label{fig1}
\end{center}
\end{figure}

The data of photoproduction of hyperon resonances are used for a
preliminary check of constituent-counting rule in
Ref.~\cite{Chang:2015ioc}. As shown in Fig.~\ref{fig1}, the best-fit
scaling factor $n$ in the region of $\sqrt{s} \ge 2.5$ GeV are 10 and
9.2 for the $\gamma \, p \to K^{+} \Lambda$ and $K^{+} \Sigma ^{0}$
reactions, respectively. This indicate that the number of the
elementary constituents is consistent with $n_{\gamma} = 1$, $n_{p} =
3$, $n_{K^{+}} = 2$, and $n_{\Lambda} = n_{\Sigma ^{0}} = 3$. Then,
the analysis is made for the photoproduction of the hyperon resonances
$\Lambda (1520)$ and $\Lambda (1405)$. The obtained values of $n$ are
9.8 and 10.6 for these two cases. The accuracy of current data is not
good enough to conclude the numbers of their quark configuration.

Similar ideas have been suggested for checking up the constituent
quark content of $\Lambda (1405)$ with pion beam at J-PARC. In
Ref.~\cite{Kawamura:2013iia} Kawamura {\it et al.} obtained the cross
section of of $\pi^- p \to K^0 \Lambda(1405)$ $d \sigma / d \Omega =
1.09 \pm 0.21 \mu b /sr$ at $\sqrt{s}=2.02$ GeV, assuming the scaling
behavior of $\Lambda(1405)$ as a 5-$q$ hadron. Then they predicted the
cross section of $\pi^- p \to K^0 \Lambda(1405)$ at CM production
angle of 90 degree to be in the order of either 100 pb or a few pb at
J-PARC energy of $\sqrt{s}=4.5$ GeV, depending on the scaling relation
of 3-$q$ or 5-$q$ nature of $\Lambda (1405)$. Because of the
relatively large $\sqrt{s}$ compared to that of photoproduction data,
the dependence of production cross sections on the nature of
constituent quark shall be good enough to be discriminated by
measurements done at J-PARC. Achieving the determination of energy
scaling behavior of $\Lambda (1405)$ shall hopefully provide a robust
experimental test of the constituent quark properties of this exotic
hadron.

\section{Exclusive Drell-Yan process}
\label{sec:excl_DY}

In recent decades, tremendous efforts have been spent in extending the
measurements to the multi-dimensional partonic structure of nucleons:
generalized parton distributions (GPDs)~\cite{Diehl:2013xca} and
transverse-momentum-dependent parton distribution functions
(TMDs)~\cite{Barone:2010zz}. The multidimensional information becomes
essential for a deeper understanding of the partonic structures of the
nucleon, including the origin of the nucleon spin and its flavor
structure. With a factorization of perturbatively calculable
short-distance hard part and universal long-distance soft hadronic
matrix elements, the nucleon parton distributions, which are the
common non-perturbative objects, could be obtained from various
spacelike and timelike reactions. Taking into account the QCD
evolution effect, the experimental verification of the universality of
the nucleon parton distributions in both spacelike and timelike
processes is essential.

\begin{figure}[htbp]
\centering
\includegraphics[width=\textwidth]{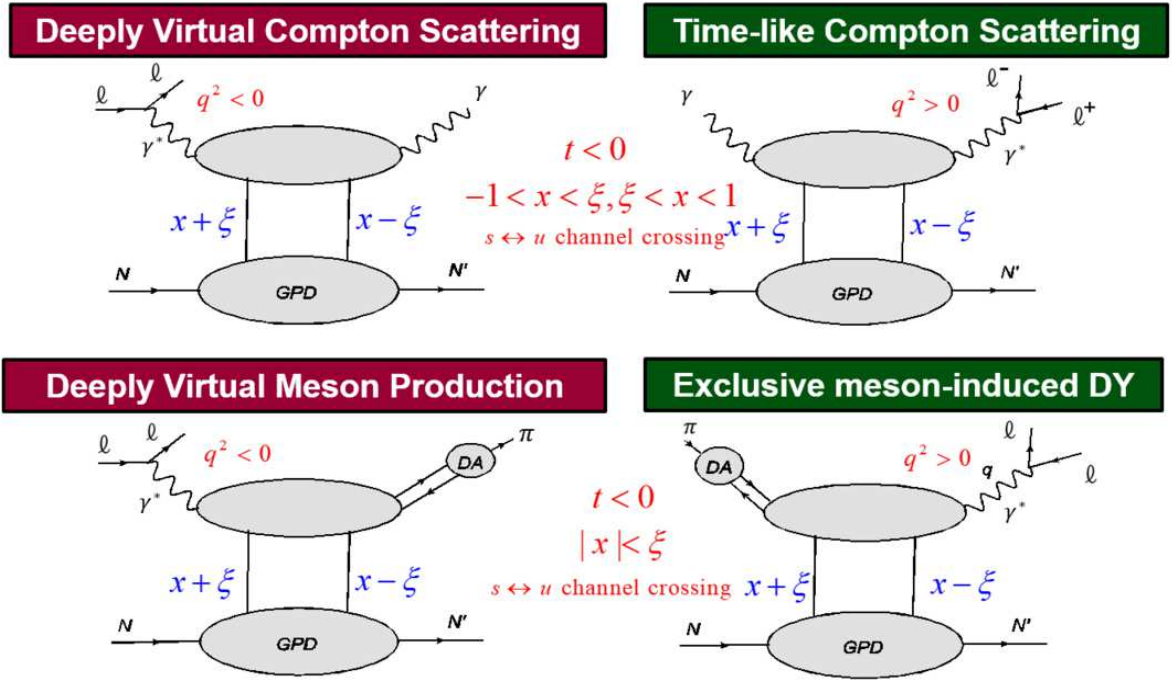}
\caption{Proposed experimental processes for accessing the nucleon
  GPDs with the lepton and pion beams.}
\label{fig:GPD}
\end{figure}

As illustrated in Fig.~\ref{fig:GPD}, GPDs were introduced in
connection with two hard exclusive processes of leptoproduction of
photons and mesons off protons: deeply virtual Compton scattering
(DVCS) and deeply virtual meson production (DVMP). There have been
many experimental activities of measuring DVCS and DVMP processes with
lepton beams. Data have been taken by HERMES, H1 and ZEUS at DESY and
HALL-A and CLAS at JLab. Recently the status of nucleon GPDs in the
valence region with the global analysis of existing DVCS and DVMP data
is reviewed in Ref.~\cite{Guidal:2013rya,Favart:2015umi}. Further
measurements were performed at COMPASS experiment at CERN and planned
for JLab after 12-GeV upgrade.

Other than lepton beams, it was suggested that GPDs could be accessed
using real photon and hadron beams as well, such as timelike Compton
scattering (TCS)~\cite{Berger:2001xd}, lepton-pair production with
meson beam~\cite{Berger:2001zn,Goloskokov:2015zsa} and pure hadronic
reaction~\cite{Kumano:2009he,Kawamura:2013wfa}. For example, invoking
the properties under time-reversal transformation and analyticity
under the change from spacelike to timelike
virtuality~\cite{Muller:2012yq}, the exclusive pion-induced Drell-Yan
process $\pi N \to \gamma^* N \to l^+ l^-
N$~\cite{Berger:2001zn,Goloskokov:2015zsa}, as illustrated in
Fig.~\ref{fig:GPD} is assumed to factorize in a way analogous to the
DVMP processes, and can serve as an independent probe to access
nucleon GPDs.

In Ref.~\cite{Berger:2001zn,Goloskokov:2015zsa}, it has been shown
that at large dilepton mass ($Q'$) scaling limit, the corresponding
leading-twist cross section of $\pi^-(q)+ {\rm p}(p) \to \gamma^*(q')+
{\rm n}(p')$ as a function of $t$ and $Q'^2$ is expressed in terms of
convolution integrals $\tilde{\cal H}^{du}$ and $\tilde{\cal E}^{du}$,
as follows~\cite{Berger:2001zn}
\begin{align}
\left.\frac{d\sigma_L}{dt dQ'^2}\right|_{\tau}
&= \frac{4\pi \alpha_{\rm em}^2}{27}\frac{\tau^2}{Q'^8} f_\pi^2\, \Bigl[ (1-\xi^2) |\tilde{\cal H}^{du}(\tilde{x},\xi,t)|^2 \nonumber \\
&- 2 \xi^2 \mbox{Re}\ \bigl( \tilde{\cal H}^{du}(\tilde{x},\xi,t)^* \tilde{\cal E}^{du}(\tilde{x},\xi,t) \bigr)
   -  \xi^2 \frac{t}{4 m_N^2}|\tilde{\cal E}^{du}(\tilde{x},\xi,t)|^2 \Bigr],
\label{eq_dcross}
\end{align}
where the scaling variable $\tilde{x}$ is given by $\tilde{x}= -
(q+q')^2 / ( 2(p+p') \cdot (q+q') ) \approx - Q'^2 / (2s - Q'^2) =
-\xi$, and the pion decay constant $f_\pi$. The subscript ``$L$'' of
the cross section indicates the contribution of the longitudinally
polarized virtual photon.

The convolution integral $\tilde{\cal H}^{du}$ involves two soft
objects: the GPD for $p\rightarrow n$ transition and the twist-two
pion distribution amplitude (DA) $\phi_{\pi}$. The expression of
$\tilde{\cal H}^{du}$ is given, at the leading order in $\alpha_s$,
by~\cite{Berger:2001zn}
\begin{align}
  \tilde{\cal H}^{du}(\tilde{x},\xi,t) &= \frac{8}{3} \alpha_s \int_{-1}^1
  dz\, \frac{\phi_\pi(z)}{1-z^2} \nonumber \\ 
  &\times \int_{-1}^1 dx
  \Bigl( \frac{e_d}{\tilde{x}-x- i\epsilon} - \frac{e_u}{\tilde{x}+x- i\epsilon}
  \Bigr) \bigl( \tilde{H}^{d}(x,\xi,t) - \tilde{H}^{u}(x,\xi,t)
  \bigr),
\label{eq_Hdu}
\end{align}
where $e_{u,d}$ are the electric charges of $u,d$ quarks in units of
the positron charge. The corresponding expression of $\tilde{\cal
  E}^{du}$ is given by (\ref{eq_Hdu}) with $\tilde{H}^q$ replaced by
the proton GPDs $\tilde{E}^q$. Due to the pseudoscalar nature of the
pion, the cross section (\ref{eq_Hdu}) receives the contributions of
$\tilde{H}$ and $\tilde{E}$ only, among the GPDs.

The leading-twist cross section (\ref{eq_dcross}) enters the four-fold
higher-twist differential cross sections for $\pi^- p \to \gamma^* n$
as~\cite{Goloskokov:2015zsa},
\begin{align}
\frac{d\sigma}{dt dQ'^2 d\cos\theta d\varphi} &= \frac{3}{8\pi} \bigl( \sin^2\theta \frac{d\sigma_L}{dt dQ'^2} + \frac{1+\cos^2\theta}{2} \frac{d\sigma_T}{dt dQ'^2} \nonumber \\ 
&+ \frac{\sin2\theta \cos\varphi}{\sqrt{2}} \frac{d\sigma_{LT}}{dt dQ'^2} + \sin^2\theta \cos2\varphi \frac{d\sigma_{TT}}{dt dQ'^2} \bigr),
\label{cadall}
\end{align}
with the angles $(\theta, \varphi)$ specifying the directions of the
decay leptons from $\gamma^*$. In Eq.~(\ref{cadall}), $d\sigma_T/(dt
dQ'^2)$ denotes the cross section contributed by the
transversely-polarized virtual photon. The $d\sigma_{LT}/(dtdQ'^2)$
and $d\sigma_{TT}/(dt dQ'^2)$ are the longitudinal-transverse
interference and transverse-transverse (between helicity $+1$ and
$-1$) interference contributions, respectively. The $d\sigma_{LT}/ (dt
dQ'^2)$ is of twist-three and is suppressed asymptotically by $1/Q'$
compared to the twist-two cross section $d\sigma_{L}/ (dt dQ'^2)$,
while $d\sigma_T/ (dt dQ'^2)$ and $d\sigma_{TT}/ (dt dQ'^2)$ are
suppressed by one more power of $1/Q'$ as twist-four effects. The
angular structures of these four terms, characteristic of the
associated virtual-photon polarizations, allow us to separate the
contribution of the leading-twist cross section $d\sigma_{L}/ (dt
dQ'^2)$ from the measured angular distributions of dilepton pairs.

\section{Feasibility study of measuring exclusive Drell-Yan process at E50 experiment}
\label{sec:E50}

Realization of the exclusive Drell-Yan measurement is interesting as
well as important to verify the universality of GPDs in both spacelike
and timelike processes. Below we present the results of feasibility
study~\cite{Sawada:2016mao} based on the known experimental beam
conditions and the setup of target and detectors of J-PARC E50
experiment with addition of a muon identification system.

The E50 experiment~\cite{E50_JPARC} plans to investigate
charmed-baryon spectroscopy via the measurement of $\pi^- + p
\rightarrow Y_{c}^{*} + D^{*-}$ reaction at the high-momentum beam
line at J-PARC. The mass spectrum of $Y_{c}^{*}$ will be constructed
by the missing-mass technique following the detection of $D^{*-}$ via
its charged decay mode. Spectroscopy of $Y_{c}^{*}$ could reveal the
essential role of diquark correlation in describing the internal
structure of hadrons. The E50 experiment has received the stage-1
approval in 2014.

\begin{figure}[hbtp]
\centering
\includegraphics[width=\textwidth]{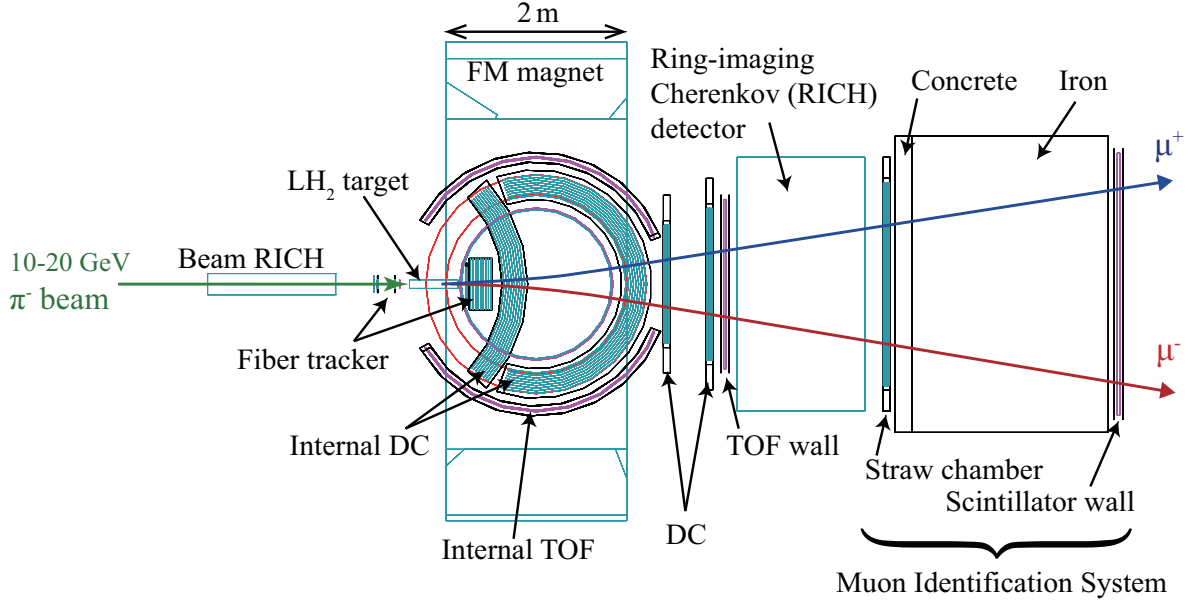}
\caption{Conceptual design of J-PARC E50 spectrometer with muon
  identification system. Figures from~\cite{Sawada:2016mao}.}
\label{fig:E50}
\end{figure}

Figure~\ref{fig:E50} shows the conceptual design of the E50
spectrometer. The spectrometer is composed of a dipole magnet and
various particle detectors~\cite{E50_JPARC}. Since the secondary beams
are unseperated, beam pions are tagged by gas Cherenkov counters (Beam
RICH) placed upstream of the target. High-granularity drift chambers
placed downstream of the magnet are for detection of charged tracks,
e.g. kaons and pions from $D^{*-}$ decay. TOF counters and
ring-imaging Cherenkov counters are placed downstream of the drift
chambers for high-momentum kaon/pion separation. In the current
spectrometer configuration, a missing-mass resolution of $D^{*-}$ is
expected to be as good as 5 MeV~\cite{E50_JPARC}.

Conventionally the measurement of Drell-Yan process in the
fixed-target experiments requires a hadron absorber immediately after
the targets to avoid large track densities in the spectrometer. Thanks
to the relatively low track density at the energy regime of J-PARC and
high-granularity tracking chambers, the measurement of Drell-Yan
process could be operated without the installation of hadron absorber
in front of the spectrometer. Excluding the multiple-scattering effect
in the hadron absorber is essential in achieving a good momentum
determination of muon tracks so that the exclusive Drell-Yan process
can be characterized via the missing-mass technique. A dedicated $\mu$
identification system, composed of tracking devices and stopping
materials, is planned to be placed in the most downstream position.

We perform the feasibility study of measuring the exclusive
pion-induced Drell-Yan process $\pi^- p \to \gamma^* n \to \mu^+\mu^-
n$ using E50 detector configuration together with $\mu$ID
system~\cite{Sawada:2016mao}. The assumed integrated luminosity is 2-4
fb$^{-1}$ for 50-day worth beam time. Both inclusive and exclusive
Drell-Yan events are generated together with the other dimuon sources
like $J/\psi$ and the random combinatorial from minimum-bias hadronic
events in the event simulation. The estimated total cross sections for
the exclusive and inclusive Drell-Yan events for the dimuon mass
$M_{\mu^{+}\mu^{-}} > 1.5$ GeV and the $|t-t_{0}|<0.5$ GeV$^2$ are
about 10-20 pb and 2-3 nb, respectively. In the range of beam momentum
10-20 GeV, the total hadronic interaction cross sections of $\pi^- p$
is about 20-30 mb while the production of $J/\psi$ is about 1-3
nb. More details are referred to Rf.~\cite{Sawada:2016mao}.

\begin{figure}[hbtp]
\centering
\hspace{-0.5cm}
\subfigure[]
{\includegraphics[width=0.33\textwidth]{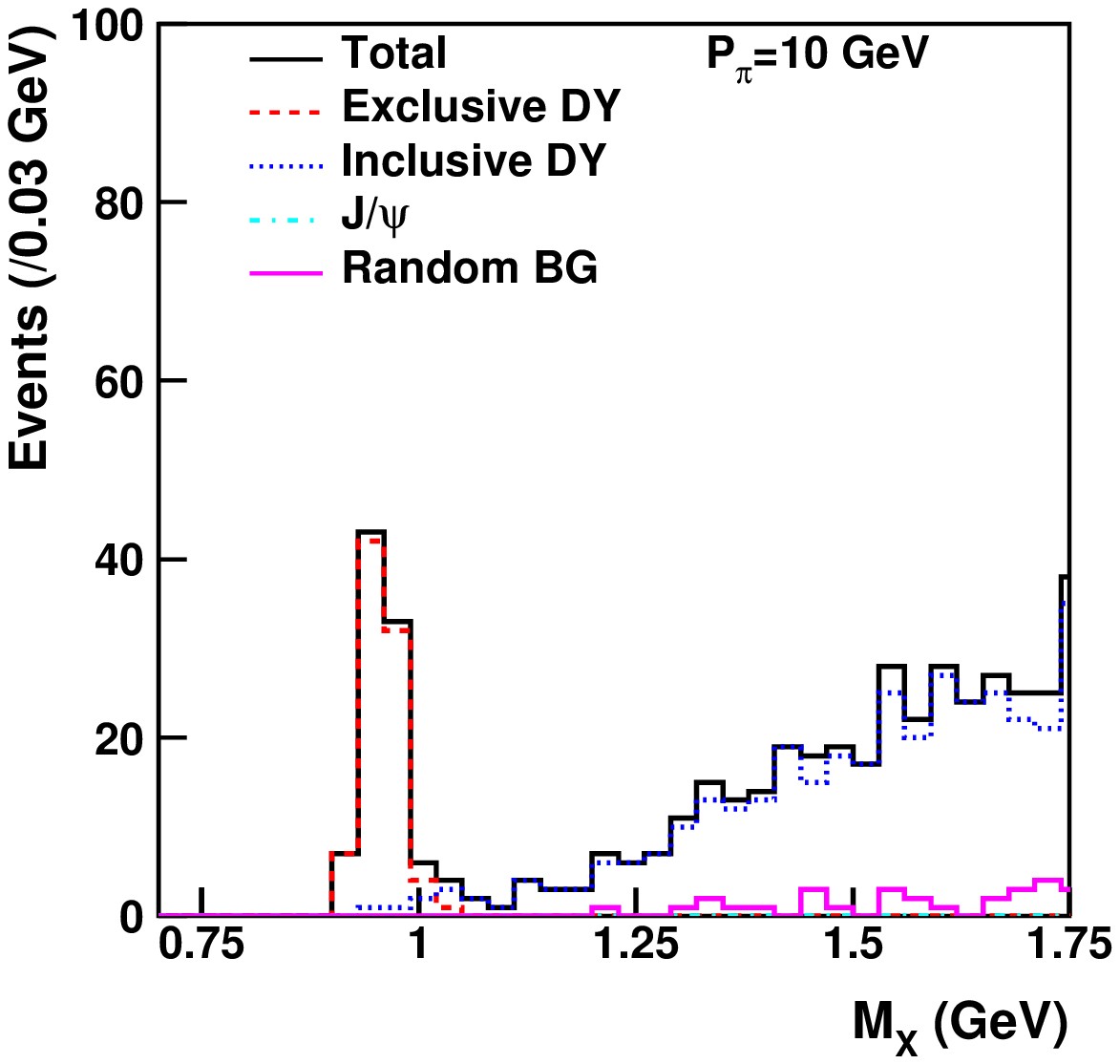}
\label{fig:mmass_1}}
\hspace{-0.5cm}
\subfigure[]
{\includegraphics[width=0.33\textwidth]{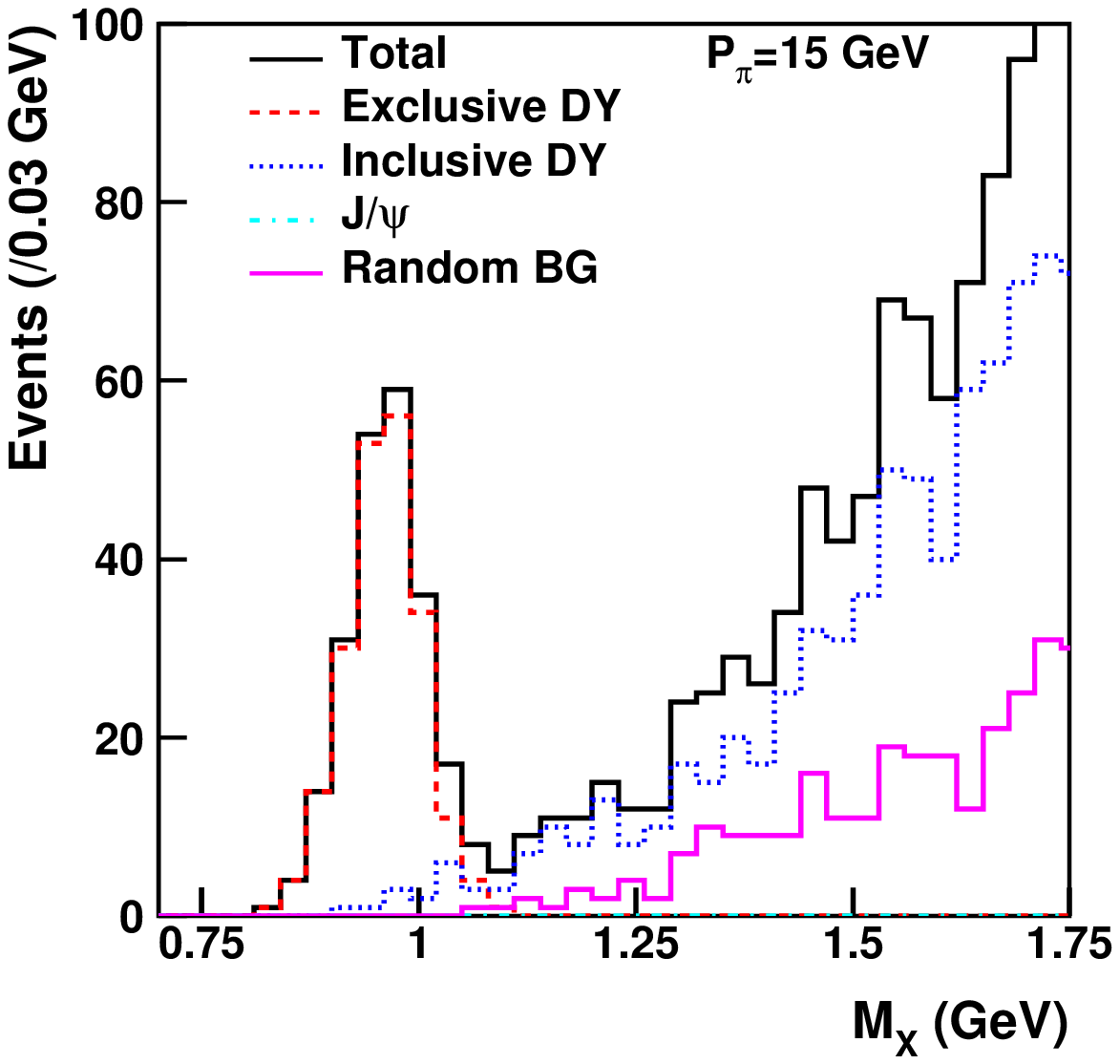}
\label{fig:mmass_2}}
\hspace{-0.5cm}
\subfigure[]
{\includegraphics[width=0.33\textwidth]{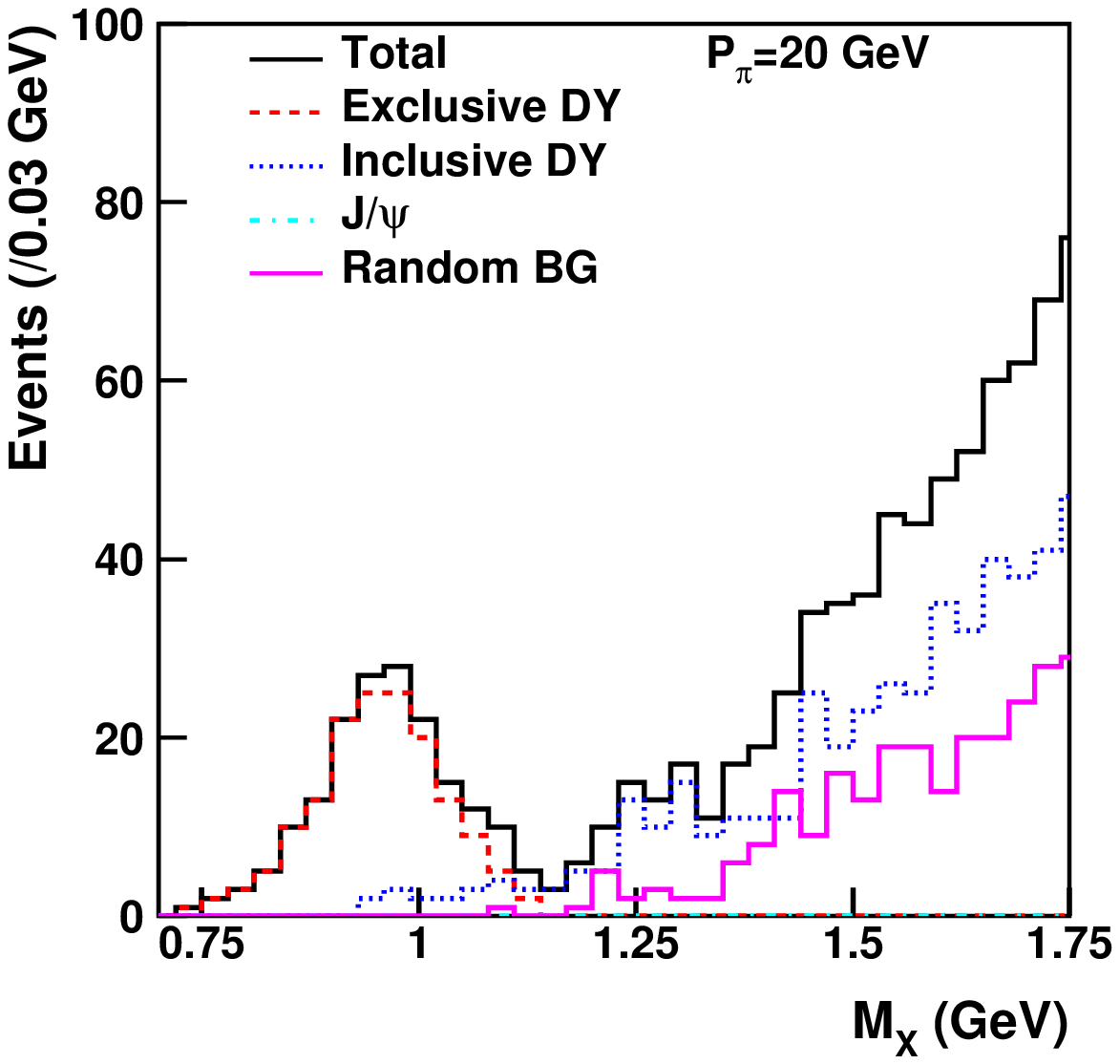}
\label{fig:mmass_3}}
\caption{The Monte-Carlo simulated missing-mass $M_{X}$ spectra of the
  $\mu^{+}\mu^{-}$ events with $M_{\mu^{+}\mu^{-}} > 1.5$ GeV and
  $|t-t_{0}|<0.5$ GeV$^2$ for $P_{\pi}$=10, 15, and 20 GeV. Lines with
  different colors denote the contributions from various sources. The
  GK2013 GPDs is used for the evaluation of exclusive Drell-Yan
  process. Figures from~\cite{Sawada:2016mao}.}
\label{fig:mmass}
\end{figure}

Using GK2013 GPDs~\cite{Kroll:2012sm} for the exclusive Drell-Yan
process, the Monte-Carlo simulated missing-mass $M_{X}$ spectra of the
$\mu^{+}\mu^{-}$ events with $M_{\mu^{+}\mu^{-}} > 1.5$ GeV and
$|t-t_{0}|<0.5$ GeV$^2$ for $P_{\pi}$=10, 15, and 20 GeV is shown in
Fig.~\ref{fig:mmass}, where $t_{0}$ is the limiting value of
4-momentum transfer square $t$.  Lines with different colors denote
the contributions from various sources: exclusive Drell-Yan (red,
dashed), inclusive Drell-Yan (blue, dotted), $J/\psi$ (cyan,
dash-dotted) and random background (purple, solid),
respectively. Signals of $J/\psi$ are only visible in the invariant
mass distributions for $P_{\pi}$=15 and 20 GeV. It is clear that the
exclusive Drell-Yan events could be identified by the signature peak
at the neutron mass ($M_n \sim 0.940$ GeV) in the missing-mass
spectrum for all three pion beam momenta. With the leading-twist
expression as Eq.~(\ref{eq_dcross}), the measured differential cross
sections $d\sigma / dt$ can be included in the global analysis for the
extraction of $\tilde{H}$ and $\tilde{E}$ GPDs. To extend the GPDs
determination to the higher-twist ones by the dimuon angular
dependence in Eq.~(\ref{cadall}), a new spectrometer with an enlarged
acceptance other than the currently designed one is preferred.

\section{Summary and Outlook}
\label{sec:summary}

Exclusive hard hadronic process is an effective way to access the
constituent quark structures. In terms of reasonable cross sections
and good enough hard scale, the meson beam of 10-20 GeV momentum to be
available at high-momentum beam line of J-PARC is most optimized for
the measurement. The energy scaling relation of $\pi^- p \to K^0
\Lambda(1405)$ can be used for exploring the constituent quark
structure of $\Lambda(1405)$, while exclusive Drell-Yan process $\pi-
p \to \gamma^* n \to \mu^+ \mu^- n$ will yield the important information
of GPDS of nucleons.

\begin{figure}[hbtp]
\centering
\includegraphics[width=0.85\textwidth]{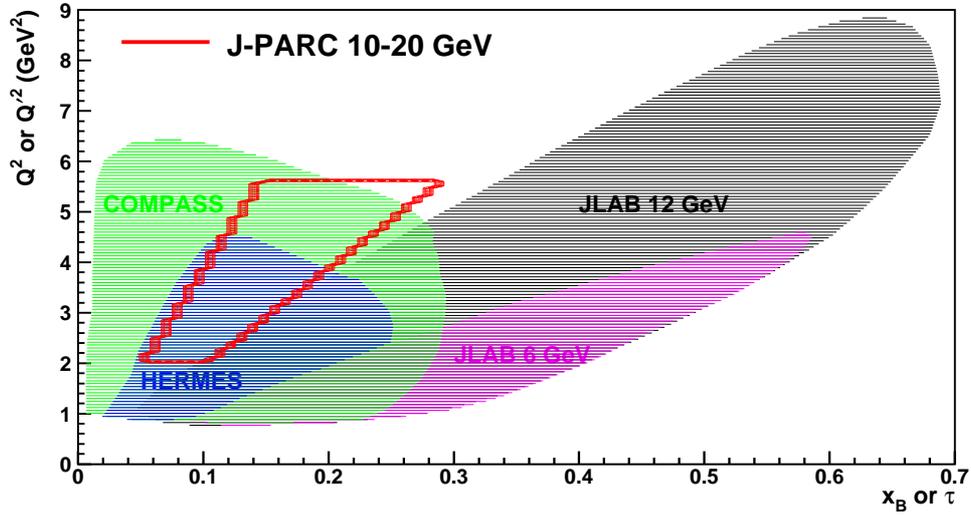}
\caption{The kinematic regions of GPDs explored by the experiments at
  JLab, HERMES and COMPASS (DVCS or DVMP) and J-PARC (exclusive Drell-Yan).
  The region is either [$Q^2$, $x_B$] for spacelike processes
  or [$Q'^2$, $\tau$] for timelike ones. Figures from~\cite{Sawada:2016mao}}
\label{fig:GPD_map}
\end{figure}

In the framework of the J-PARC E50 experiment, we addressed the
feasibility of measuring the exclusive pion-induced Drell-Yan
process. A clean signal of exclusive pion-induced Drell-Yan process
can be identified in the missing-mass spectrum of dimuon events with
2$-$4 fb$^{-1}$ integrated luminosity. Fig.~\ref{fig:GPD_map}
illustrates the kinematic regions of GPDs in terms of $Q^2$ versus
$x_B$ for spacelike processes and $Q'^2$ versus $\tau$ for timelike
ones to be explored by the existing and coming experiments. Testing
the universality of nucleon GPDs through both the measurements of
spacelike and timelike processes on the same kinematic region shall be
of fundamental importance.

A letter of intent for studying the nucleon GPDs via exclusive
Drell-Yan process at J-PARC has been submitted to the PAC of J-PARC in
2019~\cite{LoI}. With an appropriate trigger setting, the measurement
of exclusive hard process could be carried out simultaneously with the
approved program of charm-baryon spectroscopy in J-PARC E50
experiment. A full proposal to request a stage-1 approval of exclusive
Drell-Yan program is being prepared. The E50 experiment is aimed for
commissioning in 2025 as long as the secondary beam is ready.

\end{document}